
\input harvmac

\def\IR{\relax{\rm I\kern-.18em R}}

\Title{\vbox{\baselineskip12pt\hbox{RU-91-42/SCIPP 91-30}}}
{{\vbox {\centerline{Note on Discrete Gauge Anomalies}
}}}
\bigskip

\centerline{\it Tom Banks }
\medskip
\centerline{Department of Physics and Astronomy}
\centerline{Rutgers University, Piscataway, NJ 08855-0849, USA}
\medskip
\centerline{\it Michael Dine}
\centerline{Santa Cruz Institute for Particle Physics}
\centerline{University of California, Santa Cruz, CA 95064 USA}
\vskip.5in

\noindent

We consider the probem of gauging discrete symmetries. All valid
constraints on such symmetries can be understood in the
low energy theory in terms of instantons.  We note that
string perturbation theory often exhibits global discrete symmetries,
which are broken non-perturbatively.
\vskip.5in
\centerline{\it Submitted to Physical Review D, Brief Reports}
\Date{Sep. 1991}
\newsec{Introduction}

Global discrete symmetries have been considered
in particle physics in many contexts.  While no theoretical argument
convincingly rules out the existance of such symmetries,
like all global symmetries they are viewed with a certain skepticism.
Apart
from the question of how such symmetries might arise,
it is not at all clear that symmetries of this type would
survive gravitational effects such as wormholes.\ref\wormholes{S.
Hawking,Phys. Lett. {\bf 195B},337,(1987); G.V.Lavrelashvili, V.
Rubakov, and P.Tinyakov, JETP Lett. {\bf 46},167,(1987);S.Giddings and
A. Strominger,Nucl.Phys. {\bf B306},890,(1988);
 S.Coleman,Nucl. Phys. {\bf B310}
(1988) 643.}
Thus, as for continuous symmetries, it is natural to
consider the gauging of these symmetries.  Discrete gauge
symmetries were introduced into physics by Wegner in the context
of lattice theories.\ref\wegner{F.Wegner, J. Math. Phys. {\bf 12},2259,(1971)}
They appear quite frequently
in compactifications of
string theory, where they are often relics of higher dimensional
general coordinate invariance or spontaneously broken gauge
symmetries.\ref\gsw{M.B. Green, J.S. Schwarz
and E. Witten, {\it  Superstring Theory}, Cambridge University Press,
New York (1987).}  Discrete world sheet gauge symmetries play a role in
the construction of orbifolds.  In this context it is particularly clear
that discrete gauge symmetry coincides with the ancient mathematical
procedure of constructing new spaces by modding out a manifold by the
action of a discrete group.
Krauss and Wilczek \ref\krauss{L.M. Krauss and F. Wilczek,
Phys. Rev. Lett. {\bf 62} (1989) 1221.}\
pioneered the study of discrete
gauge symmetries in four dimensional physics.
They showed that
such symmetries would give hair to black holes and would be immune
to violation by quantum gravitational effects like wormholes.

More recently Ibanez and Ross \ref\ir{CERN preprints CERN-TH-6111-91
 and CERN-TH-6000-91.}\ have derived constraints on low energy
theories by requiring that all discrete symmetries be gauged.
These constraints arise because of the possibility that
the discrete symmetries may be anomalous.
Their argument involved embedding the low energy
discrete symmetry in a continuous group which is spontaneously broken at
some high energy scale.  The anomaly constraints on this continuous
symmetry, combined with the constraints on discrete charges of fermions
which gain mass upon spontaneous symmetry breakdown, give the
Ibanez Ross (IR) constraints.
IR found that applied to low energy supersymmetric models,
these constraints are quite restrictive.

Somewhat later, Preskill, Trivedi and Wise\ref\ptw{
J. Preskill, S.P. Trivedi, F. Wilczek
and M. B. Wise, Institute for Advanced
Study preprint
IASSNS-HEP-91-11 (1991).}\
pointed out that discrete gauge symmetries are constrained by the
requirement that the 't Hooft interaction induced by instantons of any
continuous gauge symmetry in the theory be invariant under the discrete
symmetry transformation.\foot{The possibility that
discrete symmetries can be broken by instantons has been
appreciated for some time; it was mentioned to one of the
present authors by E. Witten, c. 1981.  Anomalies in discrete
symmetries also been discussed by Weinberg and others in the framework
of technicolor.\ref\weinberg{S. Weinberg, Phys. Lett. {\bf 102B}
(1981) 401.}
}

The question immediately arises whether the Ibanez Ross constraints are
related to those of PTWW.  The IR constraints are stronger, but they were
derived by postulating an embedding in a particular high energy theory.
If the constraints depend
on the method of embedding, they are not useful constraints
on a low energy effective theory.  If not, one would expect to be able to
derive
them without any reference to the high energy embedding theory.
Since the only low energy constraints presently known are those which
follow from requiring that instantons of the low energy group
not break the symmetry, the IR constraints might suggest some
new low energy phenomena.
  In the
present note we will study this question.  We find that those Ibanez
Ross constraints which are nonlinear in the discrete charges\foot{We use
terminology appropriate to an abelian discrete group.  As we will see
later, the correct constraints can be stated in a way which does not
depend on the nature of the group.} can be violated in many embedding
theories.  Therefore they are not required for consistency of the low
energy theory.  Failure of these constraints at low energy implies only
that a subgroup of the full unbroken discrete gauge symmetry of the
model leaves all the low energy fields invariant.  Correspondingly, it
predicts constraints on the spectrum of certain massive \lq\lq
fractionally charged'' states.  The linear Ibanez Ross constraints
are not affected by this ambiguity.  They follow
simply from the PTWW criterion that instantons of the low energy
theory not violate the symmetry.
They are thus
required for consistency of the low energy discrete gauge
theory.\foot{In order to demonstrate this correspondence between
PTWW and IR we have had to correct a factor of two in one of the linear IR
equations, which makes the constraint somewhat stronger.}

\newsec{\bf FRACTIONAL CHARGES AND NONLINEAR DISCRETE ANOMALIES}

The IR derivation of the cubic discrete anomaly constraints is easy to
recapitulate.  Suppose for simplicity that we have a $Z_N$ discrete
symmetry in a low energy theory.  We imagine that the theory arose from
the spontaneous breakdown of a $U(1)$ gauge symmetry by a Higgs field of
charge $N$.   Assume that the ratio of any two $U(1)$ charges in the
theory is rational.  Then there is a charge $q$ (not necessarily carried
by one of the fields in the theory) such that every charge is an integer
multiple of $q$.  Normalize the $U(1)$ generator so that $q=1$.  If we
arrange the spin one half fermions in the theory into a collection of
left handed doublets, then the anomaly cancellation condition may be
written:
\eqn\sumeqn
{\sum q_L^3 = - (\sum q_i^3 + \sum {\bar{q_i}}^3 + \sum q_a^3 )}
On the left hand side of this equation we sum over the $U(1)$ charges of
all the states in the theory which are left massless after spontaneous
symmetry breakdown.  The heavy states on the right hand side are divided
into those which get Majorana masses $q_a$, and those which pair up with
another left handed field to make a Dirac mass term.  Since the mass
terms must be made gauge invariant by multiplying them by a single
valued function of the Higgs field, the charges of the heavy fields satisfy:
\eqn\consa{2q_a = 0~ {\rm mod}~ N}
\eqn\consb{q_i + \bar{q_i} = 0~ {\rm mod}~ N}
{}From this it follows that
\eqn\cubcons{\sum q_L^3 = m N + n {N^3 \over 8}}
where $m$ and $n$ are integers.  There is nothing incorrect about this
equation or its derivation.  However, it does not refer solely to
information about the low energy theory.  The integer normalization of
charges may implicitly imply things about the high energy theory in
which the light particles are embedded.  In particular, suppose that in
the above normalization all of the light particles have charges which
are multiples of an integer $L$ which divides $N$.  Then the {\it
effective} symmetry group of the low energy theory is $Z_{N\over L}$.
The anomaly constraint is nonlinear in the charges and the cubic anomaly
constraint for $Z_{N\over L}$ is not satisfied.  Similar remarks apply
to bilinear constraints involving two $Z_N$ charges and a low energy
$U(1)$ generator.\foot{As remarked by Ibanez and Ross, these constraints
are rendered uninteresting anyway by the ambiguity in normalization of
$U(1)$ charges.}

Models in which the effective symmetry group of the full
theory is larger than that of the low energy theory are rather
common.  In string theory models constructed by
``modding out" a conformal field theory by the action of a discrete
symmetry have sectors twisted under the action of the discrete
group.  These sectors need not contain any light particles (as is
the case, for example, when one mods out a Calabi-Yau space by the
action of a freely acting group).  The symmetry acting in this sector
can be larger than that of the original conformal field theory.
For example, if the original conformal theory had a $Z_N \times Z_M$
symmetry, and one mods out by the action of the $Z_N$, the twisted
sectors may exhibit a $Z_{N \times M} $ symmetry.

The nonlinear IR constraints are not totally devoid of interest.  If we
believe that a given low energy discrete symmetry must be gauged, then
their failure implies the existences of new fractionally charged states
and an enlarged symmetry group at high energy.\foot{Here we assume that
there is some scale at which we can consider the discrete symmetry to be
embedded in a four dimensional continuous gauge group.  Since we have
shown that the nonlinear constraints depend on the nature of the high
energy theory, it is not clear that their implications are the same when
the symmetry comes from geometrical considerations as in Kaluza Klein
theories.}
 However, we have not found a way to
rewrite the constraint so that it throws much light on the nature of
these states.  In general, there will be many ways to satisfy the
constraint by adding different high energy sectors to the theory.
For example, we can always make a $Z_N$ symmetry consistent by embedding
it in a $Z_{N^2}$ theory in which all the low energy fields carry
$Z_{N^2}$ charge which is equal to $0~ {\rm mod}~N$.

\newsec{\bf INSTANTONS AND DISCRETE GAUGE SYMMETRIES}

The linear IR constraints do not suffer from the difficulty
that we encountered in the previous section.  The rescaled constraints
of the $Z_N$ theory are precisely those appropriate to the low energy
$Z_{N\over L}$ theory. It is easy to see that the linear constraint
involving low energy nonabelian gauge groups is almost identical with
that of PTWW, namely that the 't Hooft
effective Lagrangian\ref\thooft{G. 't Hooft, Phys. Rev. Lett. {\bf 37} (1976);
Phys. Rev. {\bf D14} (1976) 3432.} be invariant under the discrete group.
It is perhaps worth stressing that if
this condition is not satisfied, not only is the
symmetry broken in the one-instanton sector, but gauging
the symmetry would give an inconsistent theory.
This follows from
't Hooft's argument\thooft\ that the effect of a
dilute instanton-anti-instanton gas on low momentum fermion Green's
functions in any topological sector can be summarized by insertion of this
 effective
Lagrangian.

The PTWW constraint is stronger by a factor of $2$, than that of IR, but we
can extract this extra factor from the IR method as well.  Indeed, the
source of the extra factor of $\ha$ in IR's equation is heavy Majorana
fermions.  All such fields must transform as real representations of
the low energy nonabelian gauge group.
The Dynkin index of any such representation is an even integer (in
the normalization in which the Dynkin index counts the number of fermion
zero modes in an instanton with topological charge one),
and this gives an extra factor of
two on the right hand side of the equation that precisely cancels the
$\ha$ coming from the discrete gauge charge of a Majorana field. Thus,
the corrected IR condition coincides exactly with the low energy PTWW
condition, and is valid independently of the manner in which the theory
is modified at high energy.

This derivation of the discrete anomaly constraints makes it clear that
they probe only nonperturbative gauge dynamics, a fact which is obscured
by the IR derivation.  Indeed, from the low energy point of view, any
discrete global symmetry can be gauged in perturbation theory.  It is
the dilute instanton gas which violates anomalous discrete symmetries in
weakly coupled theories.  The PTWW derivation also shows us that we
should only expect anomalies in discrete abelian groups that act by the
same phase on all fermions in the same representation of the low energy
nonabelian gauge group.  Any other transformation
can be written as such a \lq\lq flavor
blind" phase times a transformation which leaves the `t Hooft
interaction invariant.

Similar considerations apply to the linear gravitational anomaly of
discrete symmetries.  The linear IR constraint on discrete-gravitational
anomalies can be derived by noting that the minimal gravitational
instanton which is a spin manifold (so that fermion fields are well
defined), has two fermion zero modes per Weyl field.
There is one weak point in this argument for the discrete gravitational
anomaly. As for gauge instantons, the PTWW argument demonstrates the existence
 of
a problem in a particular topological sector.  In the gauge case we were
able to promote this into an argument of inconsistency for the full
theory by considering a dilute gas.  We do not know if a similar dilute
gas argument works in the gravitational
case.\ref\witten{E. Witten, Commun. Math. Phys. {\bf 100} (1985) 197.}
The mathematical
classification of four dimensional gravitational instantons which might
satisfy cluster decomposition has not yet been carried out.
 Even if we were to find such instantons, it is
not completely clear that Euclidean
considerations make sense in quantum gravity, where the action is
unbounded from below.  On the other hand, we have examined
many field theoretic and string theoretic models with gauged discrete
symmetries and have
not been able to find any which violate this condition.
We have not been able to find consistent high energy embeddings for low
energy theories which violate the linear gravitational IR constraint, as
we were able to do for the nonlinear constraints.
Thus we believe
that it is probably correct as it stands.

If we accept this argument,
 our considerations make it easy to generalize the linear IR
conditions to discrete R symmetries in supergravity.  Such symmetries
arise, for example, in Kaluza-Klein theories and string theories,
where a surviving
discrete subgroup of the higher-dimensional Lorentz symmetry
will in general transform spinors non-trivially.
To determine the linear condition, we
need only count the number of gravitino zero modes in the background
instanton field

Gauge instantons have no gravitino zero modes, while a
minimal gravitational instanton, with signature $16$ has precisely
$2$.
Thus the discrete-nonabelian gauge anomaly condition will
remain the same for R symmetries. The anomaly constraint for
the gravitational anomaly of discrete R symmetries
will be modified to
\eqn\rconsa{2\sum q_i + 2 q_{3/2} = 0~ {\rm mod}~ N}
where the sum is over all of the fermionic fields belonging
to chiral or gauge multiplets of supersymmetry, and $q_{3/2}$
is the discrete charge of the gravitino.  We emphasize that from our
point of view, the latter equation depends on an assumption about the
spectrum of zero modes in {\it allowed} gravitational instanton
backgrounds.  Since we do not have a classification of clustering
instantons in four dimensional supergravity, this analysis must be
regarded as provisional.  The true gravitational anomaly constraint will
be that the `t Hooft effective Lagrangian for quantum gravitational
instantons be invariant under the discrete symmetry that one is
proposing to gauge.

\newsec{\bf Discrete Symmetries in String Theory}

We have mentioned string theory several times to illustrate
the issues discussed in this paper.  String theory provides
numerous examples of gauged discrete symmetries.  One might try to
turn the reasoning around and ask whether discrete gauge symmetries
in string theory are ever anomalous.  Our interest here is in
string models which are free of perturbative anomalies, i.e.
modular invariant.  We know of no general argument that
insures that discrete gauge symmetries in such models are anomaly
free.  On the other hand, we have explained above that such
an anomaly would signal an inconsistency.  Thus it is possible
that there is an additional consistency condition for string
models, which is non-perturbative in nature.

We have examined a number of models for this possibility, and have
indeed found numerous examples where the linear IR conditions
are not satisfied.  However, in all of these cases, it it possible
to cancel the anomaly.  String compactifications always contain
at least one axion field, usually called the ``model-independent axion,"
which couples to the topological charge of the various gauge groups.
In all of the examples we have examined, it is possible to cancel
the anomaly by assigning a nonhomogeneous transformation law to the axion under
the discrete symmetry.  In other words, an instanton in these theories
gives rise to an expectation value for a fermionic operator, ${\cal O}$,
which is not invariant under the discrete symmetry. However, because
the axion couples to the topological charge, ${\cal O}$ is
multiplied by a factor of the form $e^{ia}$, where $a$ is the
axion field (in a suitable normalization).
If we assign a transformation law to the field $a$,
of the form $a \rightarrow a + 2 \pi q /N$ (in the case of a $Z_N$
symmetry), the full instanton amplitude is gauge invariant.
Such a non-linear transformation law means that
the gauge symmetry is spontaneously broken at a high energy
scale (of order the Planck scale).  Perturbation theory,
on the other hand, exhibits an unbroken discrete symmetry to
any finite order; this symmetry (which is not a gauge symmetry)
is explicitly broken by non-perturbative effects.  This in itself
may be phenomenologically interesting, since it suggests that it
is natural to postulate approximate discrete symmetries.

It is perhaps worthwhile to give one example of the phenomenon
we are describing.  For this, consider the $O(32)$ theory
compactified on a textbook\gsw\
example of a Calabi-Yau compactification, described by
a quintic polynomial in $CP^4$.  At a special point in the moduli
space, this model has a large discrete symmetry group, including
four $Z_5$ symmetries.  It is straightforward to check that these symmetries
all satisfy the linear IR conditions.
Now mod out this theory by
a freely-acting discrete symmetry.  In particular, ref. \gsw\ defines
a $Z_5$ symmetry called $A$.  Include also a Wilson line.
This Wilson line can be described as follows.
In the fermionic formulation of the heterotic string, there are $26$
free, left-moving fermions in this compactification.  Group them
as 6 complex fermions and 14 real ones.  If $\alpha$ is
a fifth root of unity, the Wilson line rotates three of the
complex fermions by $\alpha$, three by $\alpha^3$, and leaves
the rest untouched.  This choice is modular invariant.  It
leaves an unbroken gauge group $SU(3) \times SU(3) \times O(14)
\times U(1)^2$.  It also leaves unbroken the four original $Z_5$
symmetries.
  A straightforward calculation shows that,
for an instanton embedded in any of the three gauge groups,
the appropriate operator ${\cal O}$ transforms as $\alpha^2$
under each of the $Z_5$ symmetries.
Since the model-independent axion couples in the same way
to each of the gauge groups, letting $a \rightarrow a+6 \pi /5$
cancels the anomaly.

\newsec{\bf Conclusions}

It is sometimes argued that any discrete symmetries which might play
a role in low energy physics will be gauge symmetries.
Following Ibanez and Ross, we have considered the constraints
which must be satisfied if this is to be the case.  We have
seen that only conditions which can be derived from
low energy considerations (i.e. instantons of low energy gauge groups
and possibly gravitational instantons) hold independent of
assumptions about the high energy theory.

On the other hand, we have also provided some evidence that it
makes sense, as in the work of Preskill, Trivedi, Wilczek and Wise,
to postulate discrete symmetries which are broken only by
small, non-perturbative effects.  Indeed, we have seen that
this is a common phenomenon in string theory.
This is analogous to the situation with Peccei-Quinn symmetries.
In field theory, in both cases,
it seems somewhat unnatural to postulate the
existence of symmetries which are broken ``a little bit."
In string theory, this is a common occurrence.

Finally, we have noted that the anomaly
conditions may provide a non-perturbative constraint on
string compactifications, but we have not exhibited a modular
invariant model which fails to satisfy these conditions.  Similarly, we
are not aware of any perturbatively consistent string vacuum whose low
energy field theory suffers from a nonperturbative SU(2) anomaly.
Perhaps in string theory perturbative anomalies are the whole story.
\bigskip
\centerline{\bf Acknowledgements}

We thank Nathan Seiberg for many helpful conversations.
This work was supported in part by DOE grants DE-FG05-90ER40559.
and DE-AM03-76SF00010.

\listrefs
\end
\end